\documentclass[runningheads]{llncs}

\usepackage[T1]{fontenc}
\def\doi#1{\href{https://doi.org/\detokenize{#1}}{\url{https://doi.org/\detokenize{#1}}}}
\usepackage{graphicx}
\usepackage{listings}
\newcommand{\1}{\vspace{.1in}}

\newcommand{\bc}{\begin{center}}
	\newcommand{\ec}{\end{center}}
\newcommand{\n}{\noindent}

\usepackage{amsmath}
\usepackage{amssymb}
\usepackage{bussproofs}
\usepackage{url}

\begin{document}
	
	\title{A Cut-free, Sound and Complete Russellian Theory of Definite Descriptions\thanks{Funded by the European Union (ERC, ExtenDD, project number: 101054714). Views and opinions expressed are however those of the author(s) only and do not necessarily reflect those of the European Union or the European Research Council. Neither the European Union nor the granting authority can be held responsible for them.}}
			
	\author{Andrzej Indrzejczak\inst{1}\orcidID{0000-0003-4063-1651} \and 
	Nils K\"urbis\inst{1}\orcidID{0000-0002-3651-545}}
	\authorrunning{A. Indrzejczak, N. K\"urbis}
	%
	
	\institute{Department of Logic, University of Lodz, Poland
	\email{andrzej.indrzejczak@filhist.uni.lodz.pl} \\
	\email{nils.kurbis@filhist.uni.lodz.pl}}

	\maketitle              
	\begin{abstract}
	We present a sequent calculus for first-order logic with lambda terms and definite descriptions. The theory formalised by this calculus is essentially Russellian, but avoids some of its well known drawbacks and treats definite description as genuine terms. A constructive proof of the cut elimination theorem and a Henkin-style proof of completeness are the main results of this contribution.  
	
	
			\keywords{Definite Descriptions \and Predicate abstracts \and Sequent Calculus \and Cut Elimination}
	\end{abstract}

\section{Introduction}

Definite descriptions (DD) are complex terms commonly applied not only in natural languages but also in mathematics and computer science. In formal languages they are usually expressed by means of the iota operator, which forms terms from formulas. Thus $\imath x\varphi$ means `the (only) $x$ satisfying $\varphi$'. A DD aims to denote a unique object by virtue of a property that only it has. Sometimes a DD fails, because nothing or more than one thing has the property. A DD that succeeds to denote only one object is \emph{proper}; otherwise it is \emph{improper}. 

Definite descriptions, proper and improper, are ubiquitous not only in natural languages but also in mathematics and science (like the proper `the sum of 7 and 5' or the improper `the square root of $n$'). In formal languages the application of functional terms is the prevailing way of representing complex names. However, applying DD can outrun functional terms in many ways, since they are more expressive than functional terms, in the sense that an arbitrary functional term $f^n(t_1, \ldots, t_n)$ can be represented as a description $\imath xF^{n+1}(x, t_1, \ldots, t_n)$, where $F$ is a predicate corresponding to the function $f$. On the other hand, not every definite description, even if proper, can be expressed using functional terms; it is possible only in the case of predicates expressing functional relations, whereas every sentence can be used to form a DD. For example, both `the father of Ben' and `the daughter of Mary' may be represented as terms using the iota operator, but only the first may be represented as a functional term. Moreover, even if we can use functional terms instead of DD we enrich a language with another sort of functors in addition to predicates. This has an impact on the formalisation of valid arguments in which very often the conclusion follows on the basis of the content expressed by functional terms which is directly expressed by predicates. For example: `Adam has children' follows from `Adam is the father of Ben'. However to prove its validity, its formal representation $a = f(b) \vdash \exists xCxa$ requires two enthymematic premisses: $\forall xy(Mxy\vee Fxy\leftrightarrow Cyx)$ and $\forall xy(x = f(y) \leftrightarrow Fxy)$. Let us call the latter premiss a bridge principle allowing us to transfer information conveyed by predicates to related functions and vice versa. In general they have a form: $\forall x_1, \ldots, x_n, y(y=f^n(x_1, \ldots, x_n)\leftrightarrow F^{n+1}(y, x_1, \ldots, x_n)$ and show how the information encoded by the functional predicates is represented by predicates. In the case of using DD instead of functional terms we do not need such extra bridge principles, whereas in languages with functional terms they are necessary in an analysis of obviously valid arguments.\footnote{Some other advantages of using DD instead of functional terms are discussed in more detail in \cite{IndZaw2}.}

The usefulness of formal devices like the iota operator and other term-forming operators has recently been better recognised (cf. Tennant's \cite{Tennant2004} or Scott and Benzm\"uller's implementation of free logic using proof assistant \textit{Isabelle/HOL}~\cite{Isabelle}) also in the fields connected with computer science, like differential dynamic logic used for verification of hybrid systems \cite{DifDynLog} or description logics (see \cite{Artale2021} or \cite{Kutz2020}). Logics with DD are often implemented to enable formalisation of deep philosophical problems. e.g. Anselm's ontological argument (see the work by Oppenheimer and Zalta using the automated reasoning tool \textit{PROVER9}~\cite{Anselm} or its encoding by Blumson \cite{Blumson}). 

Since several rival theories of DD were formulated, the applicability and potential usefulness of DD was underestimated so far. 
It leads to a question which approach is the best one, at least for some specific kind of applications. In this paper we focus on the Russellian approach to definite descriptions (\cite{Russell1905} and \cite{WhitRus}) which plays a central role in this area. Although Russell's theory of DD has some controversial points, it became a standard point of reference of almost all works devoted to the
analysis of definite descriptions. Moreover, it is still widely accepted by formal logicians as a proper way of handling descriptions;
the scores of textbooks that use it as their official theory of definite descriptions count as witnesses for this claim. Russell’s theory has also strong affinities to logics closely connected with applications in constructive mathematics and computer science like
the logic of the existence predicate by Scott \cite{sco:fre00} or the definedness logic (or the logic of partial terms) of Beeson \cite{bee:fre00} and  Feferman \cite{fef:fre00}. These connections were elaborated in \cite{ind:des20}.

Russell treated DD as incomplete signs and defined their use by contextual definitions of the form: 

\bc

$\psi[x/\imath y\varphi] \ := \ \exists x(\forall y(\varphi \leftrightarrow y =x )\wedge\psi)$

\ec

\noindent but this solution leads to scoping difficulties if $\psi$ is not elementary. $\neg\psi[x/\imath y\varphi]$, e.g., is ambiguous: is the whole formula negated or only the predicate $\psi$? The method which Russell introduced in \cite{WhitRus} to draw scope distinctions is rather clumsy. Fortunately, it is possible to develop a logic which treats DD as genuine terms and yet retains desirable features of the Russellian approach. Such a logic was formalised as a natural deduction system by Kalish, Montague, and Mar~\cite{KaliMonMar80} and by Francez and Wi\k{e}ckowski \cite{FraWie2014}. These systems involve complex rules and axioms, but recently Indrzejczak \cite{Indrzejczak2021c} provided an analytic and cut-free sequent calculus equivalent to the Russellian logic as formalised in \cite{KaliMonMar80}. However, in all these systems the formal counterpart of the Russellian policy of eliminating DD from sentences must be restricted to predicate letters, which is connected with the scoping difficulties of the Russellian approach just mentioned. 

Can we offer any improvement on the state of the art? A possible strategy of avoiding these problems is to treat DD by means of a binary quantifier; this approach was formally developed by K\"urbis (cf. \cite{Kurbis2019a}, \cite{Kurbis2019b}, \cite{Kurbis2021a}, \cite{Kurbis2021b}, \cite{Kurbis2022}). However, if we want to treat DD as terms, then the introduction of the lambda operator to construct complex predicate abstracts from formulas offers a good solution. $\lambda x\varphi$ means `the property of being $\varphi$' and applied to some term, in particular to a DD, forms a formula called a lambda atom.
This device was introduced into studies of modal predicate logic by Tho\-ma\-son and Stalnaker~\cite{StalnakerThomason1968}, and the idea was further developed by Bressan \cite{Bressan} and Fitting~\cite{Fitting1975}, in particular, to  distinguish between \emph{de dicto} and \emph{de re} reading of modal operators.
Independently, this technique was used by Scales \cite{Scales} in his formulation of attributional logic, where Aristotle's distinction between the negation of a sentence and of a predicate is formally expressible. In fact, Scales seems to be the first one to apply predicate abstraction to formalise a theory of DD which relates closely to Russell's.  
Predicate abstracts were also successfully applied by Fitting and Mendelsohn~\cite{FitMen98} to obtain a theory of DD in a modal setting. This approach, with slight modifications, was further developed independently by Orlandelli \cite{Orlandelli2021} and Indrzejczak \cite{Indrzejczak2020a} to obtain cut-free sequent calculi for modal logics with DD and predicate abstracts.

In this article we focus on a different logic {\bf RL}, first introduced in \cite{IndZaw2}, which also combines the iota and lambda operators. It avoids the shortcomings of the Russellian approach while saving all its plausible features. Predicate abstracts permit us to draw scope distinctions rather more elegantly than with the Russellian scope markers and their application is more general. {\bf RL} is essentially Russellian but with DD treated as genuine terms. Nonetheless, the reductionist aspect of Russell's approach is retained in several ways. On the level of syntax the occurrences of DD are restricted to arguments of predicate abstracts to form lambda atoms. On the level of semantics DD are not defined by an interpretation function but by satisfaction clauses for lambda atoms. Eventually, on the level of calculus DD  cannot be instantiated for variables in quantifier rules but are subject to special rules for lambda atoms. This strict connection of DD with predicate abstracts avoids disadvantages of the Russellian approach connected with scoping difficulties, and, at the same time, simplifies proofs of metalogical properties.

{\bf RL} was originally characterised semantically and formalised as an analytic tableau calculus in \cite{IndZaw2}, where it was also applied for proving the Craig interpolation theorem. Here we are completing the research on {\bf RL} by providing an adequate sequent calculus for which the cut elimination theorem is proved constructively. We characterise the language, semantics and axiomatisation of {\bf RL} in section 2. Then we present the sequent calculus GRL for {\bf RL} and show its equivalence with an axiomatic Hilbert style system HRL. Section 4 contains a proof of the cut elimination theorem, and section 5 a Henkin-style proof of completeness. The paper finishes with some comparative remarks.

\section{Preliminaries}
The language $\mathcal{L}$ of {\bf RL} is standard, except that it contains the operators $\imath$ and $\lambda$. Following the remarks on the functional terms from the Introduction, as well as the original Russellian attitude towards terms, the `official' language has neither constant nor function symbols; in the completeness proof we add constants solely for the purpose of constructing models from consistent sets. As is customary in proof theoretic investigations since Gentzen, we distinguish free and bound variables graphically in deductions. It is not customary to make this distinction in semantics, and so there we won't make it either. This blend of two customs should not lead to confusion, and we are following Fitting and Mendelsohn \cite{FitMen98} in this respect. There are two disjoint sets $VAR$ of variables and $PAR$ of parameters. The former plays the role of the bound, the latter of the free variables in the presentation of the proof theory of {\bf RL}; in the presentation of the semantics, this restriction is relaxed and members of $VAR$ are permitted as free variables. The \emph{terms} of the language in the strict sense are the variables and parameters. Expressions formed by $\imath$ are admitted as terms in a more general sense: their application is restricted to predicate abstracts and they are called quasi-terms. We mention only the following formation rules for the more general notion of a formula used in the semantics: 

\begin{itemize}
	\item 
If $P^n$ is a predicate symbol (including $=$) and $t_1\ldots t_n\in VAR\cup PAR$, then $P^n(t_1, ..., t_n)$ is a formula (atomic formula). \item
If $\varphi$ is a formula, then $(\lambda x\varphi)$ is a predicate abstract. 
\item
If $\varphi$ is a formula, then $\imath x\varphi$ is a quasi-term. 
\item
If $\varphi$ is a predicate abstract and $t$ a term or quasi-term, then $\varphi t$ is a formula (lambda atom). 
\end{itemize}

\noindent $\varphi [x/t]$ denotes the result of replacing $x$ by $t$ in $\varphi$. To save space, we'll often write $\varphi_t^x$ instead of $\varphi [x/t]$. If $t$ is a variable $y$, it is assumed that $y$ is free for $x$ in $\varphi$, that is, no occurrence of $y$ becomes bound in $\varphi$ in the replacement. To save space and simplify things in the statement of semantics and in the completeness proof in section 4, we treat $\vee, \rightarrow, \exists$ as defined notions.

A \emph{model} is a structure $M=\langle D, I \rangle$, where for each $n$-argument predicate $P^n$, $I(P^n)\subseteq D^n$. An \emph{assignment} $v$ is a function $v:VAR\cup PAR\longrightarrow D$. An \emph{$x$-variant} $v'$ of $v$ agrees with $v$ on all arguments, save possibly $x$. We write $v^x_o$ to denote the $x$-variant of $v$ with $v^x_o(x) = o$.
The notion of \emph{satisfaction} of a formula $\varphi$ \emph{with} $v$, in symbols $M, v \models \varphi$, is defined as follows, where $t\in VAR\cup PAR$:

\medskip

\begin{tabular}{lcl}
	$M, v \models P^n(t_1, ..., t_n) $ & iff & $\langle v(t_1), \ldots, v(t_n) \rangle \in I (P^n)$ \\
	
	$M, v \models t_1 = t_2 $ & iff & $v(t_1) = v(t_2)$\\
	
	$M, v \models (\lambda x\psi)t$ & iff & $M, v^x_o \models \psi$, where $o = v(t)$ \\
	
	$M, v \models (\lambda x\psi)\imath y\varphi$ & iff & there is an $o\in D$ such that $M, v^x_o \models \psi$, and \\ && $M, v^x_o \models \varphi[y/x]$, and for any $y$-variant $v'$ of $v^x_o$,  
	\\ && if $M, v' \models \varphi$, then $v'(y)=o$\\
	
	$M, v \models \neg \varphi$ & iff & $M, v \not\models \varphi$, \\[.5ex]	
	$M, v \models \varphi \land \psi $ & iff & $M, v\models \varphi$ and $M, v \models \psi$,\\[.5ex]
	$M, v \models \forall x\varphi $ & iff & $M, v^x_o \models \varphi$, for all $o\in D$.
\end{tabular}

\medskip

\noindent A formula $\varphi$ is \emph{satisfiable} if there are a model $M$ and an assignment $v$ such that $M, v \models \varphi$. A formula is \emph{valid} if, for all models $M$ and assignments $v$, $M, v \models \varphi$. Semantically, HRL is identified with the set of valid formulas, {\bf RL} with the set of valid sequents. 
A set of formulas $\Gamma$ is \emph{satisfiable} iff there is some structure $M$ and an assignment $v$ such that $M$ satisfies every member of $\Gamma$ with $v$. A sequent $\Gamma\Rightarrow \Delta$ is satisfied by a structure $M$ with an assignment $v$ if and only if, if for all $\varphi\in \Gamma$, $M, v\models \varphi$, then for some $\psi\in\Delta$, $M, v \models \psi$. We symbolise this by $M, v \models\Gamma\Rightarrow \Delta$. A sequent $\Gamma\Rightarrow \Delta$ is \emph{valid} iff it is satisfied by every structure with every assignment $v$. In this case we write $\models \Gamma\Rightarrow \Delta$. 

Note that we do not characterise DD semantically by means of interpretation function $I$ as it is usually done (for example in \cite{FitMen98}, \cite{Orlandelli2021})). The syntactic restriction making DD only arguments in lambda atoms allows us to define them together as a separate satisfaction clause instead. It is closer to the original Russellian treatment of descriptions and simplifies the completeness proof.

Before presenting the sequent calculus, we briefly give the Hilbert system HRL. As we noted Russell treated DD as incomplete symbols and eliminated them by means of contextual definitions. Adopting the following axiom corresponding to his definitions
would be too simplistic: 

\medskip

$R$ \qquad $\psi(\imath y\varphi) \leftrightarrow \exists x(\forall y(\varphi \leftrightarrow y =x )\wedge\psi)$

\medskip

\noindent $R$ must be restricted to atomic $\psi$ or it is necessary to add means for marking scope distinctions. Whitehead and Russell chose the latter part, but their method is far from ideal. It is possible to avoid the problem in more elegant fashion with the help of a $\lambda$ operator. In particular, we can use it to distinguish the application of the negated predicate $\neg\psi$ to $\imath y\varphi$ from negating the application of $\psi$ to it. In the present context scoping difficulties arise only in relation to DD, and the problem is solved by restricting predication on DD to predicate abstracts. Accordingly, atomic formulas are built from predicate symbols and variables/parameters only. This is in full accordance with Russell, since the language of \emph{Principia} contains no primitive constant and function symbols: they are introduced by contextual definitions by means of DD. We modify $R$ to reflect the restriction that $\imath$ terms require $\lambda$ abstracts: 

\medskip

$R_\lambda$ \qquad $(\lambda x\psi)\imath y\varphi \leftrightarrow \exists x(\forall y(\varphi \leftrightarrow y =x )\wedge\psi)$

\medskip

\noindent This way we avoid problems with scope while permitting complex as well as primitive predicates to be applied to DD. The axiomatic system HRL for our logic {\bf RL} results from a standard axiomatization of pure first-order logic with identity and quantifier rules restricted to parameters by adding the axiom $R_\lambda$ and $\beta$-conversion for $\lambda$ but restricted again to parameters: $(\lambda x\psi)t \leftrightarrow \psi[x/t]$, where $t$ is a parameter. The adequacy of HRL will be demonstrated below.

\begin{figure}[t!]
	\centering
	\bgroup

	\n\begin{tabular}{ll}
		$(Cut)$ \ $\dfrac{\Gamma \Rightarrow \Delta, \varphi \qquad \varphi , \Pi \Rightarrow \Sigma  }{\Gamma, \Pi \Rightarrow \Delta, \Sigma }$ & $(AX)$ \ $ \varphi  \Rightarrow \varphi $
		\\[16pt]
		
		$(W\!\!\Rightarrow)$ \ $\dfrac{\Gamma \Rightarrow \Delta}{\varphi, \Gamma \Rightarrow \Delta}$
		&
		$(\Rightarrow\!\!W)$ \ $\dfrac{\Gamma \Rightarrow \Delta}{\Gamma\Rightarrow \Delta, \varphi}$
		\\[16pt]

		$(C\!\!\Rightarrow)$ \ $\dfrac{\varphi , \varphi , \Gamma \Rightarrow \Delta}{\varphi, \Gamma \Rightarrow \Delta}$
		&
		$(\Rightarrow\!\!C)$ \ $\dfrac{\Gamma \Rightarrow \Delta, \varphi , \varphi}{\Gamma \Rightarrow \Delta, \varphi}$
		\\[16pt]

		$(\neg\!\!\Rightarrow)$ \ $\dfrac{\Gamma \Rightarrow \Delta, \varphi}{\neg\varphi,
			\Gamma \Rightarrow \Delta}$ &
		$(\Rightarrow\!\!\neg)$ \ $\dfrac{\varphi, \Gamma
			\Rightarrow \Delta}{\Gamma
			\Rightarrow \Delta, \neg\varphi}$\\[16pt]
		
		$(\Rightarrow\!\!\wedge)$ \ $\dfrac{\Gamma
			\Rightarrow \Delta, \varphi \qquad
			\Gamma \Rightarrow \Delta,
			\psi}{\Gamma \Rightarrow \Delta,
			\varphi\wedge\psi}$ &
		$(\wedge\!\!\Rightarrow)$ \ $\dfrac{\varphi, \psi,
			\Gamma \Rightarrow \Delta}{\varphi\wedge\psi, \Gamma
			\Rightarrow \Delta}$\\[16pt]

		$(\vee\!\!\Rightarrow)$ \ $\dfrac{\varphi, \Gamma
			\Rightarrow \Delta \qquad \psi, \Gamma
			\Rightarrow \Delta}{\varphi\vee\psi,
			\Gamma\Rightarrow \Delta}$ &
		$(\Rightarrow\!\!\vee)$ \ $\dfrac{\Gamma
			\Rightarrow \Delta, \varphi,\psi} {\Gamma\Rightarrow \Delta,
			\varphi\vee\psi}$\\[16pt]
		
		$(\rightarrow \Rightarrow)$ \ $\dfrac{\Gamma
			\Rightarrow \Delta, \varphi \qquad\psi,
			\Gamma \Rightarrow \Delta} {\varphi\rightarrow\psi, \Gamma
			\Rightarrow \Delta}$ &
		$(\Rightarrow \rightarrow)$ \ $\dfrac{\varphi, \Gamma
			\Rightarrow \Delta, \psi}{\Gamma\Rightarrow\Delta, \varphi\rightarrow\psi}$\\[16pt]

$(\leftrightarrow\Rightarrow)$ \ $\dfrac{\Gamma \!\!\Rightarrow
		\Delta, \varphi, \psi \qquad \varphi, \psi, \Gamma \!\!\Rightarrow \Delta
		}{\varphi\!\leftrightarrow\!\psi, \Gamma \!\!\Rightarrow
		\Delta }$ & $(\forall \!\!\Rightarrow)$ \ $\dfrac{\varphi[x/b], \Gamma \!\!\Rightarrow \Delta
	}{\forall x\varphi, \Gamma \!\!\Rightarrow
	\Delta }$\\[16pt]

	$(\Rightarrow\leftrightarrow)$ \
$\dfrac{\varphi, \Gamma \!\!\Rightarrow \Delta,
		\psi \qquad \psi, \Gamma \Rightarrow \Delta, \varphi}{\Gamma \!\!\Rightarrow \Delta, 
		\varphi\!\leftrightarrow\!\psi}$  &
$(\Rightarrow\!\!\forall)$ \ $\dfrac{\Gamma \!\!\Rightarrow
		\Delta, \varphi[x/a]}{\Gamma \!\!\Rightarrow \Delta,
		\forall x\varphi}$\\[16pt]

$(\exists\!\!\Rightarrow)$ \ $\dfrac{\varphi[x/a], \Gamma
		\!\!\Rightarrow \Delta}{\exists x\varphi, \Gamma
		\!\!\Rightarrow \Delta }$ & $(\Rightarrow\!\!\exists
)$ \ $\dfrac{\Gamma \!\!\Rightarrow
		\Delta, \varphi[x/b]
		}{\Gamma \!\!\Rightarrow
		\Delta, \exists x\varphi}$ \\[16pt]

	$(=-)$ \ $\dfrac{\mbox{$\varphi[x/b_2], \Gamma \!\!\Rightarrow \Delta $}}{\mbox{$b_1 = b_2, \varphi[x/b_1], \Gamma \!\!\Rightarrow \Delta$}}$ &
	$(=+)$ \ $\dfrac{\mbox{$b=b, \Gamma \!\!\Rightarrow \Delta$}}{\mbox{$\Gamma \!\!\Rightarrow \Delta$}}$\\[16pt]
	
$(\lambda\Rightarrow)$ \ $\dfrac{\mbox{$\psi[x/b], \Gamma \!\!\Rightarrow \Delta$}}{\mbox{$(\lambda x\psi)b, \Gamma \!\!\Rightarrow \Delta$}}$ 
&
$(\Rightarrow\lambda)$ \ $\dfrac{\mbox{$\Gamma \!\!\Rightarrow \Delta, \psi[x/b] $}}{\mbox{$\Gamma \!\!\Rightarrow \Delta, (\lambda x\psi)b$}}$\\[16pt]

\end{tabular}

$(\imath_1\Rightarrow)$ \ $\dfrac{\mbox{$\varphi[y/a], \psi[x/a], \Gamma \!\!\Rightarrow \Delta$}}{\mbox{$(\lambda x\psi)\imath y\varphi, \Gamma \!\!\Rightarrow \Delta$}}$ 

\vspace{.13in}

$(\imath_2\Rightarrow)$ \ $\dfrac{\mbox{$\Gamma \!\!\Rightarrow
		\Delta, \varphi[y/b_1] \hspace{.5cm} \Gamma \!\!\Rightarrow \Delta, \varphi[y/b_2] \hspace{.5cm} b_1 = b_2, \Gamma\Rightarrow \Delta
		$}}{\mbox{$(\lambda x\psi)\imath y\varphi, \Gamma \Rightarrow
		\Delta$}}$

\vspace{.13in}

$(\Rightarrow\imath)$ \ $\dfrac{\mbox{$\Gamma \!\!\Rightarrow
		\Delta, \varphi[y/b] \hspace{.5cm} \Gamma \!\!\Rightarrow \Delta, \psi[x/b] \hspace{.5cm} \varphi[y/a], \Gamma\Rightarrow \Delta, a=b
		$}}{\mbox{$ \Gamma \Rightarrow
		\Delta, (\lambda x\psi)\imath y\varphi $}}$

\vspace{.13in}

where $a$ is a fresh parameter (Eigenvariable), not present in $\Gamma, \Delta$ and $\varphi$, whereas $b, b_1, b_2$ are arbitrary parameters. $\varphi$ in $(=-)$ is an atomic formula.

\egroup	
\caption{Calculus GRL}
\label{fig::Calculus}
\end{figure}

\section{Sequent Calculus}
We now formalise the Russellian logic {\bf RL} as a sequent calculus GRL. Sequents $\Gamma\Rightarrow\Delta$ are ordered pairs of finite multisets of formulas, called the antecedent and the succedent, respectively. 
GRL is essentially the calculus G1c of Troelstra and Schwichtenberg \cite{tro:bas96} with rules for identity and lambda atoms: see Fig. 1.

Let us recall that formulas displayed in the schemata are active, whereas the remaining ones are parametric, or form a context. In particular, all active formulas in the premisses are called side formulas, and the one in the conclusion is the principal formula of the respective rule application. 
Proofs are defined in the standard way as finite trees with nodes labelled by sequents. The height of a proof ${\cal D}$ of $\Gamma\Rightarrow\Delta$ is defined as the number of nodes of the longest branch in ${\cal D}$. $\vdash_k \Gamma\Rightarrow\Delta$ means that $\Gamma\Rightarrow\Delta$ has a proof with height at most $k$. $\vdash$ means that there is a proof of the expression standing to its right, be it a formula (in the case of HRL) or a sequent (in the case of GRL).  

We need some auxiliary results. In particular, since $(=-)$ is Leibniz' Principle restricted to atomic formulas, we must prove its unrestricted form.

\begin{lemma}\label{leibnizandsubstitution}
	\begin{enumerate}
		\item
		$\vdash b_1=b_2, \varphi[x/b_1] \Rightarrow\varphi[x/b_2]$, for any formula $\varphi$.
		\item
		If $\vdash_k \Gamma \Rightarrow \Delta$, then $\vdash_k \Gamma[b_1/b_2] \Rightarrow \Delta[b_1/b_2]$, where $k$ is the height of a proof.  
	\end{enumerate}	
\end{lemma}

\noindent \begin{proof} 1. follows by induction over the complexity of formulas, which is standard for all cases except those concerning lambda atoms with DD. We note that $\varphi{^z_b}{^y_c}$ is the same as $\varphi{^y_c}{^z_b}$, etc. We write $[(\lambda x\psi)\imath y\varphi]_{b_1}^z$ to denote substitutions in lambda atoms in more readable fashion. To simplify proofs applications of weakening and contraction rules to derive shared contexts are omitted from now on. Let $\mathcal{D}$ be the following deduction, where the leaves are axioms and $c$ a fresh parameter: 
 
\begin{footnotesize}
\begin{prooftree}
\AxiomC{$\varphi{^y_c}{^z_{b_1}} \Rightarrow \varphi{^y_c}{^z_{b_1}}$}
\AxiomC{$\varphi{^y_a}{^z_{b_1}} \Rightarrow \varphi{^y_a}{^z_{b_1}}$}
\AxiomC{$c=a\Rightarrow c=a$}
\LeftLabel{$(\imath_2\Rightarrow)$}
\TrinaryInfC{$[(\lambda x\psi)\imath y\varphi]^z_{b_1}, \varphi{^y_a}{^z_{b_1}}, \varphi{^y_c}{^z_{b_1}} \Rightarrow c = a$}
\end{prooftree}
\end{footnotesize}

\noindent Then we derive $\vdash b_1=b_2, [(\lambda x\psi)\imath y\varphi]^z_{b_1} \Rightarrow [(\lambda x\psi)\imath y\varphi]^z_{b_2}$: 

\begin{footnotesize}
\begin{prooftree}
\AxiomC{$b_1 = b_2, \varphi{^y_a}{^z_{b_1}} \Rightarrow \varphi{^y_a}{^z_{b_2}}$}
\AxiomC{$b_1 = b_2, \psi{^x_a}{^z_{b_1}} \Rightarrow \psi{^x_a}{^z_{b_2}}$}
\AxiomC{${\cal D}$}
 \LeftLabel{$(\Rightarrow\imath)$}
 \TrinaryInfC{$b_1 = b_2, \varphi{^y_a}{^z_{b_1}}, \psi{^x_a}{^z_{b_1}}, [(\lambda x\psi)\imath y\varphi]^z_{b_1} \Rightarrow [(\lambda x\psi)\imath y\varphi]^z_{b_2}$}
\LeftLabel{$(\imath_1\Rightarrow)$}
\UnaryInfC{$b_1 = b_2, [(\lambda x\psi)\imath y\varphi]^z_{b_1}, [(\lambda x\psi)\imath y\varphi]^z_{b_1} \Rightarrow [(\lambda x\psi)\imath y\varphi]^z_{b_2}$}
\LeftLabel{$(C\Rightarrow)$}
\UnaryInfC{$b_1 = b_2, [(\lambda x\psi)\imath y\varphi]^z_{b_1} \Rightarrow [(\lambda x\psi)\imath y\varphi]^z_{b_2}$}
 \end{prooftree}
 \end{footnotesize}
 
\noindent The two left leaves are provable by the induction hypothesis (if $b_1, b_2$ are not present in $\psi$ or $\varphi$, we have an axiomatic sequent). 

The proof of 2 is by a standard induction on the height of proofs; the rules for lambda atoms with DD are treated similarly to the rules for quantifiers. \qed \end{proof}

\medskip

Let us now show that the Russellian axiom $R_\lambda$ is provable in GRL. We will provide proofs for two sequents corresponding to two implications. Let $\mathcal{D}$ be: 

\begin{footnotesize}
\begin{prooftree}
\AxiomC{$\varphi{^y_a} \Rightarrow \varphi{^y_a}$}
\AxiomC{$\varphi{^y_{a_1}} \Rightarrow \varphi{^y_{a_1}}$}
\AxiomC{$a_1=a\Rightarrow a_1=a$}
\LeftLabel{$(\imath_2\Rightarrow)$}
\TrinaryInfC{$(\lambda x\psi)\imath y\varphi, \varphi{^y_a}, \varphi{^y_{a_1}} \Rightarrow a_1 = a$}
\end{prooftree}
\end{footnotesize}

\noindent The following establishes one half of $R_\lambda$:

\begin{footnotesize}
\begin{prooftree}
\AxiomC{$\mathcal{D}$}
\AxiomC{$\varphi{^y_a}, a_1=a \Rightarrow \varphi{^y_{a_1}}$}
\LeftLabel{$(\Rightarrow\leftrightarrow)$}
\BinaryInfC{$(\lambda x\psi)\imath y\varphi, \varphi{^y_a}\Rightarrow \varphi{^y_{a_1}} \leftrightarrow a_1 = a$}
\LeftLabel{$(\Rightarrow\forall)$}
\UnaryInfC{$(\lambda x\psi)\imath y\varphi, \varphi{^y_a}\Rightarrow \forall y(\varphi \leftrightarrow y = a)$}
\AxiomC{$\psi{^x_a}\Rightarrow \psi{^x_{a}}$}
\LeftLabel{$(\Rightarrow\wedge)$}
\BinaryInfC{$(\lambda x\psi)\imath y\varphi, \psi^x_a, \varphi{^y_a}\Rightarrow \forall y(\varphi \leftrightarrow y = a)\wedge\psi^x_a$}
\LeftLabel{$(\Rightarrow\exists)$}
\UnaryInfC{$(\lambda x\psi)\imath y\varphi, \psi^x_a, \varphi{^y_a}\Rightarrow \exists x(\forall y(\varphi \leftrightarrow y = x)\wedge\psi)$}
\LeftLabel{$(\imath_1\Rightarrow)$}
\UnaryInfC{$(\lambda x\psi)\imath y\varphi, (\lambda x\psi)\imath y\varphi\Rightarrow \exists x(\forall y(\varphi \leftrightarrow y = x)\wedge\psi)$}
\LeftLabel{$(C\Rightarrow)$}
\UnaryInfC{$(\lambda x\psi)\imath y\varphi\Rightarrow \exists x(\forall y(\varphi \leftrightarrow y = x)\wedge\psi)$}
\end{prooftree}
\end{footnotesize}

\noindent where the only nonaxiomatic sequent is provable by lemma \ref{leibnizandsubstitution}.1.
Next, where $\mathcal{D}$ is:

\begin{footnotesize}
\begin{prooftree}
\AxiomC{$\varphi^y_b\Rightarrow \varphi^y_b$}
\AxiomC{$b=a\Rightarrow b=a$}
\LeftLabel{$(\leftrightarrow\Rightarrow)$}
\BinaryInfC{$\varphi^y_b\leftrightarrow b=a, \varphi^y_b\Rightarrow b=a$}
\LeftLabel{$(\forall\Rightarrow)$}
\UnaryInfC{$\forall y(\varphi\leftrightarrow y=a), \varphi^y_b\Rightarrow b=a$}
\end{prooftree}
\end{footnotesize}

\noindent the following establishes the other half of $R_\lambda$:

\begin{footnotesize}
\begin{prooftree}
\AxiomC{$\psi^x_a \Rightarrow \psi^x_a$}
\AxiomC{$a=a\Rightarrow a=a$}
\LeftLabel{$(=+)$}
\UnaryInfC{$\Rightarrow a=a$}
\AxiomC{$\varphi^y_a\Rightarrow \varphi^y_a$}
\LeftLabel{$(\leftrightarrow\Rightarrow)$}
\BinaryInfC{$\varphi^y_a\leftrightarrow a=a\Rightarrow \varphi^y_a$}
\LeftLabel{$(\forall\Rightarrow)$}
\UnaryInfC{$\forall y(\varphi\leftrightarrow y=a)\Rightarrow \varphi^y_a$}
\AxiomC{$\mathcal{D}$}
\LeftLabel{$(\Rightarrow\imath)$}
\TrinaryInfC{$\forall y(\varphi\leftrightarrow y=a), \psi^x_a\Rightarrow (\lambda x\psi)\imath y\varphi$}
\LeftLabel{$(\wedge\Rightarrow)$}
\UnaryInfC{$\forall y(\varphi\leftrightarrow y=a)\wedge\psi^x_a\Rightarrow (\lambda x\psi)\imath y\varphi$}
\LeftLabel{$(\exists\Rightarrow)$}
\UnaryInfC{$\exists x(\forall y(\varphi\leftrightarrow y=x)\wedge\psi)\Rightarrow (\lambda x\psi)\imath y\varphi$}
\end{prooftree}
\end{footnotesize}

Conversely, the three rules for lambda atoms with DD are derivable in G1 with $R_\lambda$ added in the form of two axiomatic sequents. To derive $(\imath_1\Rightarrow)$, let $R_\lambda^\Rightarrow$ be $(\lambda x\psi)\imath y\varphi\Rightarrow \exists x(\forall y(\varphi \leftrightarrow y = x)\wedge\psi)$: 

\begin{footnotesize}
\begin{prooftree}
\AxiomC{$R_\lambda^\Rightarrow$}
\AxiomC{$a=a\Rightarrow a=a$}
\LeftLabel{$(=+)$}
\UnaryInfC{$\Rightarrow a=a$}
\AxiomC{$\varphi^y_a, \psi^x_a, \Gamma\Rightarrow \Delta$}
\LeftLabel{$(\leftrightarrow\Rightarrow)$}
\BinaryInfC{$\varphi^y_a\leftrightarrow a=a, \psi^x_a, \Gamma\Rightarrow \Delta$}
\LeftLabel{$(\forall\Rightarrow)$}
\UnaryInfC{$\forall y(\varphi\leftrightarrow y=a), \psi^x_a, \Gamma\Rightarrow \Delta$}
\LeftLabel{$(\wedge\Rightarrow)$}
\UnaryInfC{$\forall y(\varphi\leftrightarrow y=a)\wedge\psi^x_a, \Gamma\Rightarrow \Delta$}
\LeftLabel{$(\exists\Rightarrow)$}
\UnaryInfC{$\exists x(\forall y(\varphi \leftrightarrow y = x)\wedge\psi), \Gamma\Rightarrow \Delta$}
\LeftLabel{$(Cut)$}
\BinaryInfC{$(\lambda x\psi)\imath y\varphi, \Gamma\Rightarrow\Delta$}
\end{prooftree}
\end{footnotesize}

\noindent To derive $(\imath_2\Rightarrow)$, use $(Cut)$ with $(\lambda x\psi)\imath y\varphi\Rightarrow \exists x(\forall y(\varphi \leftrightarrow y = x)\wedge\psi)$ and:  

\begin{footnotesize}
\begin{prooftree}
\AxiomC{$\Gamma\Rightarrow \Delta, \varphi^y_{b_1}$}
\AxiomC{$\Gamma\Rightarrow \Delta, \varphi^y_{b_2}$}
\AxiomC{$b_1=b_2, \Gamma\Rightarrow \Delta$}
\LeftLabel{$(=-)$}
\UnaryInfC{$b_1=a, b_2=a, \Gamma\Rightarrow \Delta$}
\LeftLabel{$(\leftrightarrow\Rightarrow)$}
\BinaryInfC{$b_1=a, \varphi^y_{b_2}\leftrightarrow b_2=a, \Gamma\Rightarrow \Delta$}
\LeftLabel{$(\leftrightarrow\Rightarrow)$}
\BinaryInfC{$\varphi^y_{b_1}\leftrightarrow b_1=a, \varphi^y_{b_2}\leftrightarrow b_2=a, \Gamma\Rightarrow \Delta$}
\LeftLabel{$(\forall\Rightarrow)$}
\UnaryInfC{$\forall y(\varphi\leftrightarrow y=a), \forall y(\varphi\leftrightarrow y=a), \Gamma\Rightarrow \Delta$}
\LeftLabel{$(C\Rightarrow)$}
\UnaryInfC{$\forall y(\varphi\leftrightarrow y=a), \psi^x_a, \Gamma\Rightarrow \Delta$}
\LeftLabel{$(\wedge\Rightarrow)$}
\UnaryInfC{$\forall y(\varphi\leftrightarrow y=a)\wedge\psi^x_a, \Gamma\Rightarrow \Delta$}
\LeftLabel{$(\exists\Rightarrow)$}
\UnaryInfC{$\exists x(\forall y(\varphi \leftrightarrow y = x)\wedge\psi), \Gamma\Rightarrow \Delta$}
\end{prooftree}
\end{footnotesize}

\noindent The following derives $(\Rightarrow\imath)$: 

\begin{footnotesize}
\begin{prooftree}
\AxiomC{$\varphi^y_a, \Gamma\Rightarrow \Delta, a=b$}
\AxiomC{$\Gamma\Rightarrow \Delta, \varphi^y_b$}
\AxiomC{$a=b, \varphi^y_b\Rightarrow \varphi^y_a$}
\LeftLabel{$(Cut)$}
\BinaryInfC{$a=b, \Gamma\Rightarrow \Delta, \varphi^y_a$}
\LeftLabel{$(\Rightarrow\leftrightarrow)$}
\BinaryInfC{$\Gamma\Rightarrow\Delta, \varphi^y_a\leftrightarrow a=b$}
\LeftLabel{$(\Rightarrow\forall)$}
\UnaryInfC{$\Gamma\Rightarrow\Delta, \forall y(\varphi\leftrightarrow y=b)$}
\AxiomC{$\Gamma \Rightarrow \Delta, \psi^x_b$}
\LeftLabel{$(\Rightarrow\wedge)$}
\BinaryInfC{$\Gamma\Rightarrow\Delta, \forall y(\varphi\leftrightarrow y=b)\wedge\psi^x_b$}
\LeftLabel{$(\Rightarrow\exists)$}
\UnaryInfC{$\Gamma\Rightarrow\Delta, \exists x(\forall y(\varphi\leftrightarrow y=x)\wedge\psi)$}
\end{prooftree}
\end{footnotesize}

\noindent  where the right premiss of $(Cut)$ is provable by lemma \ref{leibnizandsubstitution}.1, and the conclusion of the rule follows by $(Cut)$ with $\exists x(\forall y(\varphi\leftrightarrow y=x)\wedge\psi)\Rightarrow (\lambda x\psi)\imath y\varphi$.

Since the proofs of the interderivability of the axiom of $\lambda$ conversion and $(\lambda\Rightarrow), (\Rightarrow\lambda)$ are trivial we are done and conclude with:
  
  \begin{theorem}
  	$\vdash_{HRL}\varphi$ iff \ $\vdash_{GRL} \ \Rightarrow\varphi$
  \end{theorem}

\section{Cut Elimination}
We will show that $(Cut)$ is eliminable from every proof in GRL using the general strategy of cut elimination proofs applied originally for hypersequent calculi in Metcalfe, Olivetti and Gabbay \cite{gab:pro12}, which works well also in the context of standard sequent calculi (see \cite{ind:sat}). Such a proof has a particularly simple structure and allows us to avoid many complexities inherent in other methods of proving cut elimination. In particular, we avoid well known problems with contraction, since two auxiliary lemmata deal with this problem in advance. We assume that all proofs are regular in the sense that every parameter $a$ which is fresh by the side condition of the respective rule must be fresh in the entire proof, not only on the branch where the application of this rule takes place. There is no loss of generality since every proof may be systematically transformed into a regular one by lemma \ref{leibnizandsubstitution}.2. The following notions are crucial for the proof:

\begin{enumerate}
	\item
	The cut-degree is the complexity of the cut-formula $\varphi$, i.e. the number of logical constants (connectives, quantifiers and operators) occurring in $\varphi$; it is denoted by $d\varphi$.
	\item
	The proof-degree ($d{\cal D}$) is the maximal cut-degree in ${\cal D}$.
\end{enumerate}

\noindent The proof of the cut elimination theorem is based on two lemmata which successively make a reduction: first of the height of the right, and then of the height of the left premiss of cut. $\varphi^k, \Gamma^k$ denote $k > 0$ occurrences of $\varphi, \Gamma$, respectively. 

\begin{lemma}[Right reduction]
	Let ${\cal D}_1 \vdash \Gamma\Rightarrow\Delta, \varphi$ and ${\cal D}_2 \vdash \varphi^k, \Pi \Rightarrow\Sigma$ with
	$d{\cal D}_1, d{\cal D}_2 < d\varphi$, and $\varphi $ principal in $\Gamma\Rightarrow\Delta, \varphi$, 
	then we can construct a proof ${\cal D}$ such
	that ${\cal D} \vdash \Gamma^k, \Pi \Rightarrow \Delta^k, \Sigma$ and $d{\cal D} < d\varphi$.
\end{lemma}

\begin{proof}
	By induction on the height of ${\cal D}_2$. The basis is trivial, since $\Gamma\Rightarrow\Delta, \varphi$ is identical with $\Gamma^k, \Pi \Rightarrow \Delta^k, \Sigma$.
	The induction step
	requires examination of all cases of possible derivations of $\varphi^k, \Pi \Rightarrow\Sigma$,
	and the role of the cut-formula in the transition. In cases where all occurrences of $\varphi $ are
	parametric we simply apply the induction hypothesis to the premisses of $\varphi^k, \Pi \Rightarrow\Sigma$ and then
	apply the respective rule -- it is essentially due to the context independence of almost all rules and the regularity of
	proofs, which together prevent violation of side conditions on eigenvariables. 
	If one of the occurrences of $\varphi $ in the premiss(es) is a side formula of the last rule we
	must additionally apply weakening to restore the missing formula before the application of the relevant rule. 
	
	In cases where one occurrence of
	$\varphi $ in $\varphi^k, \Pi \Rightarrow\Sigma$ is principal we make use of the fact that $\varphi $ in the
	left premiss is also principal; for the cases of contraction and weakening this is trivial. We consider the cases of lambda atoms with DD. Hence ${\cal D}_1$ finishes with:
	
	\1
	
	$\dfrac{\mbox{$\Gamma \!\!\Rightarrow
			\Delta, \varphi[y/b] \hspace{.5cm} \Gamma \!\!\Rightarrow \Delta, \psi[x/b] \hspace{.5cm} \varphi[y/a], \Gamma\Rightarrow \Delta, a=b
			$}}{\mbox{$ \Gamma \Rightarrow
			\Delta, (\lambda x\psi)\imath y\varphi $}}$
		
		\1
	
	\noindent and ${\cal D}_2$ finishes with:
	
	\1
	
	$\dfrac{\mbox{$\varphi[y/a'], \psi[x/a'], (\lambda x\psi)\imath y\varphi^{k-1}, \Pi \!\!\Rightarrow \Sigma$}}{\mbox{$(\lambda x\psi)\imath y\varphi^k, \Pi \!\!\Rightarrow \Sigma$}}$ 
	
	\1
	
	\noindent or
	
	\1
	
\n	\begin{scriptsize}
	$\dfrac{\mbox{$(\lambda x\psi)\imath y\varphi^{k-1}, \Pi \!\!\Rightarrow
			\Sigma, \varphi[y/b_1] \hspace{.5cm} (\lambda x\psi)\imath y\varphi^{k-1}, \Pi \!\!\Rightarrow \Sigma, \varphi[y/b_2] \hspace{.5cm} b_1 = b_2, (\lambda x\psi)\imath y\varphi^{k-1}, \Pi\Rightarrow \Sigma
			$}}{\mbox{$(\lambda x\psi)\imath y\varphi^k, \Pi \Rightarrow
			\Sigma$}}$
	\end{scriptsize}
	
	\1
	
In the first case, by the induction hypothesis and lemma \ref{leibnizandsubstitution}.2    we obtain	
$\varphi[y/b], \psi[x/b], \Gamma^{k-1}, \Pi \!\!\Rightarrow \Delta^{k-1}, \Sigma$	and by two cuts with the leftmost and central premiss of $(\Rightarrow\imath)$ in ${\cal D}_1$ we obtain  $\Gamma^{k+1}, \Pi \!\!\Rightarrow \Delta^{k+1}, \Sigma$, which by contraction yields the result.

In the second case note first that by lemma  \ref{leibnizandsubstitution}.2  from the rightmost 
	premiss of $(\Rightarrow\imath)$ in ${\cal D}_1$ we obtain 
	
	\medskip
	
	a. $\varphi[y/b_1], \Gamma\Rightarrow \Delta, b_1=b$ and 
	
	b. $\varphi[y/b_2], \Gamma\Rightarrow \Delta, b_2=b$. 
	
	\medskip
	
	\noindent Again by the induction hypothesis from the three premisses we get:
	
	\medskip
	
	1. $\Gamma^{k-1}, \Pi \!\!\Rightarrow \Delta^{k-1}, 
	\Sigma, \varphi[y/b_1]$
	
	2.  $\Gamma^{k-1}, \Pi \!\!\Rightarrow \Delta^{k-1}, 
	\Sigma, \varphi[y/b_2]$
	
	3. $b_1=b_2, \Gamma^{k-1}, \Pi \!\!\Rightarrow \Delta^{k-1}, 
	\Sigma$
	
	\medskip

\noindent We proceed as follows with a series of the applications of cut, followed by contractions, using the provable sequent $b_1=b, b_2=b \Rightarrow b_1=b_2$:

\begin{scriptsize}	
\begin{prooftree}
\AxiomC{$2$}
\AxiomC{$b$}
\BinaryInfC{$\Gamma^{k}, \Pi \!\!\Rightarrow \Delta^{k}, \Sigma, b_2=b$}
\AxiomC{$1$}
\AxiomC{$a$}
\BinaryInfC{$\Gamma^{k}, \Pi \!\!\Rightarrow \Delta^{k}, \Sigma, b_1=b$}
\AxiomC{$b_1=b, b_2=b \Rightarrow b_1=b_2$}
\AxiomC{$3$}
\BinaryInfC{$b_1=b, b_2=b, \Gamma^{k-1}, \Pi \!\!\Rightarrow \Delta^{k-1}, \Sigma$}
\BinaryInfC{$b_2=b, \Gamma^{2k-1}, \Pi^2 \!\!\Rightarrow \Delta^{2k-1}, \Sigma^2$}
\BinaryInfC{$\Gamma^{3k-1}, \Pi^3 \!\!\Rightarrow \Delta^{3k-1}, \Sigma^3$}
\UnaryInfC{$\Gamma^{k}, \Pi \!\!\Rightarrow \Delta^{k}, \Sigma$}
\end{prooftree}
\end{scriptsize}
\qed \end{proof}

\begin{lemma}[Left reduction]\label{leftred}
	Let ${\cal D}_1 \vdash \Gamma\Rightarrow\Delta, \varphi^k$ and ${\cal D}_2 \vdash \varphi, \Pi \Rightarrow\Sigma$ with
	$d{\cal D}_1, d{\cal D}_2 < d\varphi$,  
	then we can construct a proof ${\cal D}$ such
	that ${\cal D} \vdash \Gamma, \Pi^k \Rightarrow \Delta, \Sigma^k$ and $d{\cal D} < d\varphi$.
\end{lemma}

\begin{proof} By induction on the height of ${\cal D}_1$ but with some important differences to the proof of the right reduction lemma. 
	First note that we do not require $\varphi $ to be principal in $\varphi, \Pi \Rightarrow\Sigma$, so it includes the case where 
	$\varphi$ is atomic. In all these cases we just apply the 
	induction hypothesis. This guarantees that even if an atomic cut formula was introduced in the right premiss by $(=-)$ the reduction of the height is achieved only on
	the left premiss, and we always obtain the expected result. 
	Now, in cases where one occurrence of
	$\varphi $ in $\Gamma\Rightarrow\Delta, \varphi^k$ is principal, we first apply the induction hypothesis to eliminate all other $k-1$ occurrences of $\varphi$ in the premisses and then we apply
	the respective rule. Since the only new occurrence of $\varphi$ is principal, we can make use of the right reduction lemma again and obtain the result, possibly after some applications of structural rules. \qed \end{proof}

Now we are ready to prove the cut elimination theorem: 

\begin{theorem}
	Every proof in GRL can be transformed into cut-free proof. 
\end{theorem}

\begin{proof} By double induction: primary on $d{\cal D}$ and subsidiary on the number of maximal cuts (in the basis and
	in the inductive step of the primary induction). 
	We always take the topmost maximal cut and apply lemma \ref{leftred} to it.
	By successive repetition of this procedure we reduce either the degree of a proof or the number of
	cuts in it until we obtain a cut-free proof. \qed \end{proof}

\section{Adequacy}
In this section, we'll make use of the fact that for every set there is a corresponding multiset, so if $\Gamma$, $\Delta$ are sets of formulas, we may write $\Gamma\Rightarrow \Delta$. We recall that we treat $\vee, \rightarrow, \exists$ as defined notions. For the completeness proof we assume that a denumerable set of individual constants may be added to the language. $I$ assigns objects in the domain $D$ of the model $\langle D, I\rangle$ to these constants. For brevity we introduce the notation $I_v$, where if $t$ is a variable or parameter, $I_v(t)=v(t)$ and where $t$ is a constant, $I_v(t)=I(t)$. 

Recall the distinction between terms and pseudo-terms, the former variables and parameters and now also constants, the latter iota terms. In the following lemma, $t$ denotes a variable, parameter or constant, not a DD,  hence the proof is standard, with the case of lambda atoms similar to the case of quantifiers. In the rest of this section, too, $t$ will refer to terms only. In particular, there is no need to consider pseudo-terms in the Lindenbaum-Henkin construction (theorem \ref{maxconset}), because in substitution in the formulas concerned only terms can be used. Pseudo-terms are treated, just as they are in the semantics, as occurring in lambda atoms, and thus like the logical constants by the consideration of the consistent addition of formulas to a set in the construction of its maximally consistent extension. 

\begin{lemma}[The Substitution Lemma.]
	$M, v \models \varphi^x_t$ iff $M, v^x_{I_v(t)} \models \varphi$, if $t$ is free for $x$ in $\varphi$. 
\end{lemma} 

\begin{proof} See e.g. \cite[133f]{endertonlogic} and adjust.  \qed \end{proof}

\noindent Next, the soundness of GRL. 

\begin{theorem}[Soundness of GRL]\label{Theorem3}
	If $\vdash \Gamma \Rightarrow \Delta$, then $\models \Gamma \Rightarrow \Delta$
	\end{theorem}

\begin{proof} 
By induction on the height of the proof. Since it is well-known that the rules of G1 are validity preserving, and it is obvious for both lambda rules, we show this property only for $(\imath_2\Rightarrow)$ and $(\Rightarrow \imath)$, leaving $(\imath_1\Rightarrow)$ as an exercise. 
		
	\medskip
	
	\noindent $(\imath_2\Rightarrow)$. Suppose (1) $\models \Gamma \Rightarrow \Delta, \varphi_{b_1}^y$, (2) $\models \Gamma\Rightarrow\Delta, \varphi_{b_2}^y$, (3) $\models b_1=b_2, \Gamma\Rightarrow \Delta$, and $\not\models (\lambda x\psi)\imath y\varphi, \Gamma \Rightarrow \Delta$. By the last, there are a structure $M=\langle D, I\rangle$ and assignment $v$, such that $M, v\models (\lambda x\psi)\imath y\varphi$, for all $\gamma\in\Gamma$, $M, v\models \gamma$ and for all $\delta\in\Delta$, $M, v\not\models \delta$. Thus by (1), (2) and (3): (4) $M, v\models \varphi_{b_1}^y$, (5) $M, v\models \varphi_{b_2}^y$ and (6) $M, v\not\models b_1=b_2$. And there is an $o\in D$ such that $M, v^x_o \models \psi$, and $M, v^x_o \models \varphi[y/x]$, and (7) for any $y$-variant $v'$ of $v^x_o$, if $M, v' \models \varphi$, then $v'(y)=o$. By the conventions on the use of free and bound variables in sequents, $x$ is not free in $\varphi_{b_1}^y$ or $\varphi_{b_2}^y$, so $v$ and $v_o^x$ agree on them, and so by (4) and (5) $M, v_o^x\models \varphi_{b_1}^y$ and $M, v_o^x\models \varphi_{b_2}^y$. By the substitution lemma, $M, v{_o^x}{^y_{I_v({b_1})}}\models \varphi$ and $M, v{_o^x}{^y_{I_v({b_2})}}\models \varphi$. So the $y$-variants $v'$ and $v''$ of $v_o^x$ that assign $I_{v_o^x}(b_1)$ and $I_{v_o^x}(b_2)$ to $y$ satisfy $\varphi$ with $M$, so by (7) $I_{v'}(b_1)=I_{v''}(b_2)=o$. But $v'$ and $v''$ differ from $v$ only in what they assign to $x$ and $y$, and by (6) $I_v(b_1)\not=I_v(b_2)$. Contradiction. 
	
	\medskip
	
\noindent $(\Rightarrow \imath)$. Suppose (1) $\models \Gamma\Rightarrow \Delta, \varphi_b^y$, (2) $\models \Gamma\Rightarrow \Delta, \psi_b^x$, (3) $\models \varphi_a^y, \Gamma\Rightarrow \Delta, a=b$, but $\not\models\Gamma\Rightarrow \Delta, (\lambda x\psi)\imath y\varphi$, $a$ not free in any formulas in $\Gamma$ and $\Delta$ nor in $\varphi$. Then there are a structure $M=\langle D, I\rangle$ and assignment $v$ such that for all $\gamma\in\Gamma$, $M, v\models \gamma$, for all $\delta\in\Delta$, $M, v\not\models \delta$ and (4) $M, v\not\models (\lambda x\psi)\imath y\varphi$. So by (1), $M, v\models \varphi_b^y$, by (2), $M, v\models\psi_b^x$, and by (4), it is not the case that there is an $o\in D$ such that $M, v^x_o \models \psi$, and $M, v^x_o \models \varphi_x^y$, and for any $y$-variant $v'$ of $v^x_o$, if $M, v' \models \varphi$, then $v'(y)=o$, i.e. for every $o\in D$, either $M, v^x_o \not\models \psi$, or $M, v_o^x\not\models \varphi_x^y$, or for some $y$-variant $v'$ of $v^x_o$, $M, v' \models \varphi$ and $v'(y)\not =o$. Consider $I_v(b)$. We have either (5) $M, v^x_{I_v(b)} \not\models \psi$, or (6) $M, v_{I_v(b)}^x\not\models \varphi_x^y$, or (7) for some $y$-variant $v'$ of $v^x_{I_v(b)}$, $M, v' \models \varphi$ and $v'(y)\not ={I_v(b)}$. By the substitution lemma from (5) and (6) we have $M, v\not\models \psi_b^x$ and $M, v\not\models \varphi{_y^x}{^y_b}$, and as $\varphi{_x^y}{^x_b}$ is the same as $\varphi^y_b$, this contradicts consequences of (1) and (2). By conventions on the use of free and bound variables in sequents, $x$ and $y$ are not free in any of their formulas, so $v^x_{I_v(b)}$ agrees with $v$ on all formulas in $\Gamma$, $\Delta$, so for all $\gamma\in\Gamma$, $M, v^x_{I_v(b)}\models \gamma$, and for all $\delta\in\Delta$, $M, v^x_{I_v(b)}\not\models \delta$. So by (3), if $M, v^x_{I_v(b)}\models \varphi_a^y$, then $M, v^x_{I_v(b)}\models a=b$. By the substitution lemma and the semantic clause for identity, if $M, v{^x_{I_v(b)}}{^y_{I_v(a)}}\models \varphi$, then $I_v(a)=I_v(b)$. Now evidently $v{^x_{I_v(b)}}{^y_{I_v(a)}} (y)=I_v(a)$, so $v{^x_{I_v(b)}}{^y_{I_v(a)}} (y)=I_v(b)$. But $v{^x_{I_v(b)}}{^y_{I_v(a)}}$ is a $y$-variant of $v{^x_{I_v(b)}}$, and the reasoning holds for any such $y$-variant, contradicting (7). \qed \end{proof}


\medskip

\noindent Let $\bot$ represent an arbitrary contradiction. A set of formulas $\Gamma$ is \emph{inconsistent} iff $\Gamma\vdash \bot$. $\Gamma$ is \emph{consistent} iff it is not inconsistent. A set of formulas $\Gamma$ is \emph{maximal} iff for any formula $A$, either $A\in \Gamma$ or $\neg A\in \Gamma$. A set of formulas $\Gamma$ is \emph{deductively closed} iff, if $\Gamma\vdash A$, then $A\in\Gamma$. We state without proof this standard result:

\begin{lemma}
Any maximally consistent set is deductively closed. 
\end{lemma}



\noindent Extend $\mathcal{L}$ to a language $\mathcal{L}^+$ by adding countably new constants ordered by a list $\mathcal{C}=c_1, c_2\ldots$. We will say that such a constant occurs \emph{parametrically} if its occurrence satisfies the restrictions imposed on parameters in $(\Rightarrow\forall)$ and $(\imath_1\Rightarrow)$. 

\begin{theorem}\label{maxconset}
	Any consistent set  of formulas $\Delta$ can be extended to a maximally consistent set $\Delta^+$ such that: 
	
	\noindent (a) for any formula $\varphi$ and variable $x$, if $\neg\forall x\varphi\in \Delta^+$, then for some constant $c$, $\varphi_c^x\not\in\Delta^+$; 
	
	\noindent (b) for any formulas $\varphi$, $\psi$ and variables $x$, $y$, if $(\lambda x\psi)\imath y\varphi\in\Delta^+$, then for some constant $c$, $\varphi_c^y, \psi_c^x\in\Delta^+$ and for all terms $t$, if $\varphi_t^y\in\Delta^+$, then $t=c\in\Delta^+$; 
	
	\noindent (c) for any formulas $\varphi$, $\psi$ and variables $x$, $y$, if $\neg (\lambda x\psi)\imath y\varphi\in\Delta^+$, then for all terms $t$, either $\varphi_t^y\not\in\Delta^+$, or for some constant $c$, $\varphi_c^y\in\Delta^+$ and $c=t\not\in\Delta^+$, or $\psi_t^x\not\in\Delta^+$. 
\end{theorem}

\begin{proof} Extend $\Delta$ by following an enumeration $\phi_1, \phi_2\ldots$ of the formulas of $\mathcal{L}^+$ on which every formula occurs infinitely many times as follows:\medskip

$\Delta_0=\Delta$

\noindent If $\Delta_n, \phi_n$ is inconsistent, then 

$\Delta_{n+1} = \Delta_n$. 

\noindent If $\Delta_n, \phi_n$ is consistent, then: 

\noindent (i) If $\phi_n$ has neither the form $\neg\forall x\varphi$ nor $(\lambda x\psi)\imath y\varphi$ nor $\neg (\lambda x\psi)\imath y\varphi$, then 

$\Delta_{n+1}=\Delta_n, \phi_n$.

\noindent (ii) If $\phi_n$ has the form $\neg\forall x\varphi$, then 

$\Delta_{n+1}=\Delta_n, \neg\forall x\varphi, \neg \varphi_c^x$ 

\noindent where $c$ is the first constant of $\mathcal{C}$ that does not occur in $\Delta_n$ or $\phi_n$. 

\noindent (iii) If $\phi_n$ has the form $(\lambda x\psi)\imath y\varphi$, then 

$\Delta_{n+1}=\Delta_n, (\lambda x\psi)\imath y\varphi, \varphi_c^y, \psi_c^x$

\noindent where $c$ is the first constant of $\mathcal{C}$ that does not occur in $\Delta_n$ or $\phi_n$.  

\noindent (iv) If $\phi_n$ has the form $\neg (\lambda x\psi)\imath y\varphi$, then

$\Delta_{n+1}=\Delta_n, \neg (\lambda x\psi)\imath y \varphi, \Sigma_n$

\noindent where $\Sigma_n$ is constructed in the following way. Take a sequence of formulas $\sigma_1, \sigma_2\ldots$ of the form $\varphi_t^y\rightarrow ( \psi_t^x\rightarrow \neg (\varphi_c^y\rightarrow c=t))$, where $t$ is a term in $\Delta_n, \phi_n$, and $c$ is a constant of $\mathcal{C}$ not in $\Delta_n, \phi_n$ or any previous formulas in the sequence. Let $\mathcal{T}=t_1, t_2, \ldots$ be an enumeration of all terms occurring in $\Delta_n, \phi_n$. In case $\Delta_0$ contains infinitely many formulas, it must be ensured that $\mathcal{C}$ is not depleted of constants needed later. So pick constants from $\mathcal{C}$ by a method that ensures some constants are always left over for later use. The following will do. Let $\sigma_1$ be $\varphi_{t_1}^y\rightarrow (\psi_{t_1}^x\rightarrow \neg (\varphi_{c_1}^y\rightarrow c_1=t_1))$, where $t_1$ is the first term of $\mathcal{T}$ and $c_1$ is the first constant of $\mathcal{C}$ not in $\Delta_n, \phi_n$; let $\sigma_2$ be $\varphi_{t_2}^y\rightarrow (\psi_{t_2}^x\rightarrow \neg (\varphi_{c_2}^y\rightarrow c_2=t_2))$, where $t_2$ is the second term on $\mathcal{T}$ and $c_2$ is the $2^2=4$th constant of $\mathcal{C}$ not in $\Delta_n, \phi_n, \sigma_1$. In general, let $\sigma_n$ be $\varphi_{t_n}^y\rightarrow (\psi_{t_n}^x\rightarrow \neg (\varphi_{c_n}^y\rightarrow c_n=t_n))$, where $t_n$ is the $n$th term of $\mathcal{T}$ and $c_n$ is the $2^n$th constant of $\mathcal{C}$ not in $\Delta_n, \phi_n$ nor any $\sigma_i$, $i<n$. The entire collection of $\sigma_i$s is $\Sigma_n$.\medskip

\noindent $\Delta_{n+1}$ is consistent if $\Delta_n, \phi_n$ is:\medskip

\noindent Case (i). Trivial.\medskip

\noindent Case (ii). Suppose $\Delta_{n+1}=\Delta_n, \neg\forall x\varphi, \neg \varphi_c^x$ is inconsistent. Then for some finite $\Delta_n'\subseteq\Delta_n$: $\vdash \Delta_n', \neg\forall x\varphi, \neg \varphi_c^x\Rightarrow\bot$. Hence $\vdash \Delta_n', \neg\forall x\varphi  \Rightarrow \varphi_c^x$ by deductive properties of negation. $c$ does not occur in any formula in $\Delta_n'$ nor in $\neg\forall x\varphi$, so it occurs parametrically, and so by $(\Rightarrow\forall)$, $\vdash\Delta_n', \neg\forall x\varphi\Rightarrow\forall x\varphi$. Hence $\vdash\Delta_n' \Rightarrow\forall x\varphi$, again by deductive properties of negation. But then $\Delta_n', \neg\forall x\varphi$ is inconsistent, and hence so is $\Delta_n, \neg\forall x\varphi$.\medskip

\noindent Case (iii). Suppose $\Delta_{n+1}=\Delta_n, (\lambda x\psi)\imath y\varphi, \varphi_c^y, \psi_c^x$ is inconsistent. Then for some finite $\Delta_n'\subseteq\Delta_n$, $\vdash\Delta_n', (\lambda x\psi)\imath y\varphi, \varphi_c^y, \psi_c^x\Rightarrow\bot$. $c$ does not occur in $\Delta_n',  (\lambda x\psi)\imath y\varphi$, so it occurs parametrically, and hence by $(\imath_1\Rightarrow)$, $\vdash\Delta_n',  (\lambda x\psi)\imath y\varphi\Rightarrow\bot$, that is to say $\Delta_n',  (\lambda x\psi)\imath y\varphi$ is inconsistent, and so is $\Delta_n, (\lambda x\psi)\imath y\varphi$.\medskip 

\noindent Case (iv). Suppose $\Delta_{n+1}=\Delta_n, \neg (\lambda x\psi)\imath y\varphi, \Sigma_n$ is inconsistent. Then for some finite $\Delta_n'\subseteq\Delta_n$ and a finite $\{\sigma_j\ldots \sigma_k\}\subseteq\Sigma_n$, $\vdash\Delta_n', \neg (\lambda x\psi)\imath y\varphi, \sigma_j\ldots \sigma_k\Rightarrow\bot$. Let $\sigma_k$ be $\varphi_{t_k}^y\rightarrow (\psi_{t_k}^x\rightarrow \neg (\varphi_{c_k}^y\rightarrow \ c_k=t_k))$. Then by the deductive properties of implication and negation: 

$\vdash\Delta_n', \neg (\lambda x\psi)\imath y\varphi, \sigma_j\ldots \sigma_{k-1}\Rightarrow \varphi_{t_k}^y$

$\vdash\Delta_n', \neg (\lambda x\psi)\imath y\varphi, \sigma_j\ldots \sigma_{k-1}\Rightarrow \psi_{t_k}^x$ 

$\vdash\Delta_n', \neg (\lambda x\psi)\imath y\varphi, \sigma_j\ldots \sigma_{k-1}, \varphi_{c_k}^y\Rightarrow c_k=t_k$

\noindent $c_k$ was chosen so as not to occur in any previous $\sigma_i$, $i<k$, nor in $\Delta_n, \phi_n$. Hence it occurs parametrically and the conditions for $(\Rightarrow\imath)$ are fulfilled. Thus $\vdash\Delta_n', \neg (\lambda x\psi)\imath y\varphi, \sigma_j\ldots \sigma_{k-1}\Rightarrow (\lambda x\psi)\imath y\varphi$. But $\vdash\Delta_n', \neg (\lambda x\psi)\imath y\varphi, \sigma_j\ldots \sigma_{k-1}\Rightarrow \neg (\lambda x\psi)\imath y\varphi$. So $\Delta_n', \neg (\lambda x\psi)\imath y\varphi, \sigma_j\ldots \sigma_{k-1}$ is inconsistent. Repeat this process from $\sigma_{k-1}$ all the way down to $\sigma_j$, showing that $\Delta_n', \neg (\lambda x\psi)\imath y\varphi$ is inconsistent. Hence so is $\Delta_n, \neg (\lambda x\psi)\imath y\varphi$.\medskip 

\noindent Let $\Delta^+$ be the union of all $\Delta_i$. $\Delta^+$ is maximal, for if neither $\varphi$ not $\neg \varphi$ are in $\Delta^+$, then there is a $\Delta_k\subseteq\Delta^+$ such that $\Delta_k, \varphi\vdash\bot$ and $\Delta_k, \neg \varphi\vdash\bot$, but then $\Delta_k$ is inconsistent, contradicting the method of construction of $\Delta_k$. $\Delta^+$ is consistent, because otherwise some $\Delta_i$ would have to be inconsistent, but they are not. 

$\Delta^+$ satisfies (a) by construction. 

To see that it satisfies (b), suppose $(\lambda x\psi)\imath y\varphi\in\Delta^+$. Then there is a $\Delta_{n+1}=\Delta_n, (\lambda x\psi)\imath y\varphi, \varphi_c^y, \psi_c^x$, and so $\varphi_c^y, \psi_c^x\in\Delta^+$. Suppose $\varphi_t^y\in\Delta^+$. Then there is a $\Delta'\subseteq\Delta^+$ such that $\vdash\Delta'\Rightarrow \varphi_c^y$, $\vdash\Delta'\Rightarrow \varphi_t^y$ and by properties of identity $\vdash t=c\Rightarrow t=c$. But then by $(\imath_2\Rightarrow)$, $\vdash\Delta', (\lambda x\psi)\imath y \varphi\Rightarrow t=c$, hence $t=c\in\Delta^+$ by the deductive closure of $\Delta^+$. 

To see that it satisfies (c), suppose $\neg (\lambda x\psi)\imath y\varphi\in\Delta^+$, but for some term $t$, $\varphi_t^y\in\Delta^+$, (1) for all constants $c$, if $\varphi_c^y\in\Delta^+$, then $c=t\in\Delta^+$, and $\psi_t^x\in\Delta^+$. As every formula occurs infinitely many times on the enumeration of formulas of $\mathcal{L}^+$, there is a $\Delta_n$ that contains $\varphi_t^y$ and $\psi_t^x$ and $\Delta_{n+1}=\Delta_n, \neg (\lambda x\psi)\imath y\varphi, \Sigma_n$. Thus $\varphi_t^y\rightarrow (\psi_t^x \rightarrow \neg (\varphi_b^y\rightarrow b=t))\in\Sigma_n$, for some constant $b$ of $\mathcal{C}$. Consequently, this formula is in $\Delta^+$, too. By the deductive properties of implication and negation and the deductive closure and consistency of $\Delta^+$, (2) $\varphi_b^y\in\Delta^+$ and $b=t\not\in\Delta^+$. But by (1) and (2),  $b=t\in\Delta^+$. Contradiction. 

This completes the proof of Theorem \ref{maxconset}. \qed \end{proof}

\begin{theorem}\label{satisfiability}
	If $\Delta$ is a consistent set of formulas, then $\Delta$  is satisfiable. 
\end{theorem}

\begin{proof} Extend $\Delta$ to a maximally consistent set $\Delta^+$ as per Theorem \ref{maxconset}. We construct a structure $M=\langle D, I\rangle$ and function $v\colon VAR\cup PAR\to D$ from $\Delta^+$ which will satisfy $\Delta$. $D$ is the set of equivalence classes of terms under identities $t_1=t_2\in\Delta^+$. Denote the equivalence class to which $t$ belongs by $[t]$. For all predicate letters $P$, $\langle [t_1], ..., [t_n]\rangle \in I(P^n)$ iff $P^n(t_1, ..., t_n)\in\Delta^+$. For all variables $v(x)=[x]$, and for all parameters $v(a)=[a]$. In these latter cases $I_v=v$, and for all new constants of $\mathcal{C}$, $I_v(c)=[c]$. We'll show by induction over the number of logical constants (connectives, quantifiers, $\imath$ and $\lambda$ symbols) in formula $\varphi$ that $M, v\models \varphi$ if and only if $\varphi\in\Delta^+$.\medskip

\noindent Suppose $\varphi$ is an atomic formula. (a) $\varphi$ is $P^n(t_1, ..., t_n)$. Then $M, v\models P^n(t_1, ..., t_n)$ iff $\langle I_v(t_1), ..., I_v(t_n)\rangle\in I(P^n)$, iff $\langle [t_1]\ldots [t_n]\rangle\in I(P^n)$, iff $P^n(t_1, ..., t_n)\in\Delta^+$. (b) $\varphi$ is $t_1=t_2$. Then $M, v\models t_1=t_2$ iff $I_v(t_1)=I_v(t_2)$, iff $[t_1]=[t_2]$, and as these are equivalence classes under identities in $\Delta^+$, iff $t_1=t_2\in\Delta^+$. 

\medskip

\noindent For the rest of the proof suppose $M, v\models \varphi$ if and only if $\varphi\in\Delta$, where $\varphi$ has fewer than $n$ connectives. We skip the standard cases of $\neg, \land, \forall$ (see e.g. \cite{endertonlogic}). \medskip






\noindent Case 4. $\varphi$ is $(\lambda x\psi) t$. 

\noindent $(\lambda x\psi)t\in \Delta^+$ iff $\psi_t^x\in\Delta^+$ by deductive closure of $\Delta^+$, iff $M, v\models \psi_t^x$ by induction hypothesis. $t$ must be free for $x$ in $\psi$, hence by the substitution lemma, $M, v\models \psi_t^x$ iff $M, v_{I_v(t)}^x \models \psi$, iff $M, v_{[t]}^x \models \psi$ and $I_v(t)=[t]$, as the latter holds by construction of $M$, and this in turn is the case iff $M, v\models (\lambda x\psi) t$ by the first semantic clause for lambda atoms.\medskip

\noindent Case 5. $\varphi$ is $(\lambda x\psi)\imath y\chi$. 

\noindent (a) If $(\lambda x\psi)\imath y\chi\not\in\Delta^+$, then by deductive closure $\neg (\lambda x\psi)\imath y\chi\in\Delta^+$, and so for all terms $t$, either $\chi_t^y\not\in\Delta^+$, or for some constant $c$, $\chi_c^y\in\Delta^+$ and $c=t\not\in\Delta^+$, or $\psi_t^x\not\in\Delta^+$. $[t]\in D$ iff $t$ is a term, so by induction hypothesis, for all  $[t]\in D$, either $M, v\nvDash \chi_t^y$, or there is a $[c]\in D$ such that $M, v\models \chi_c^y$ and $M, v\nvDash c=t$, or $M, v\nvDash \psi_t^x$. $\chi_t^y$ is the same formula as $\chi{_x^y}{_t^x}$, 
so $M, v\nvDash \chi{_x^y}{_t^x}$. Furthermore, $x$ and $y$ are not free in $\chi_c^y$, so for any $o\in D$, $M, v\models \chi_c^y$ iff $M, v_o^x\models \chi_c^y$. 
By the substitution lemma, either $M, v_{I_v(t)}^x\nvDash \chi_x^y$, or $M, v_{I_v(t)}^x\nvDash \psi$, or there is a $[c]\in D$ such that $M, v_{I_v(t)}^x{_{I_v(c)}^y}\models \chi$ and $M, v_{I_v(t)}^x{_{I_v(c)}^y}\nvDash y=x$. $I_v(t)=[t]$ and $I_v(c)=[c]$, so either $M, v_{[t]}^x\nvDash \chi_x^y$, or $M, v_{[t]}^x\nvDash \psi$, or there is a $[c]\in D$ such that $M, v_{[t]}^x{_{[c]}^y}\models \chi$ and $M, v_{[t]}^x{_{[c]}^y}\nvDash y=x$, i.e. $v_{[t]}^x{_{[c]}^y}(y)\not=[t]$. $v_{[t]}^x{_{[c]}^y}$ is a $y$-variant of $v_{[t]}^x$, hence $M, v\nvDash (\lambda x\psi)\imath y\chi$. 

\noindent (b) If $(\lambda x\psi)\imath y\chi\in\Delta^+$, then for some constant $c$, $\psi_c^x, \chi_c^y \in\Delta^+$ and for all terms $t$, if $\chi_t^y\in\Delta^+$, then $c=t\in\Delta^+$. By induction hypothesis, $M, v\models \psi_c^x$ and $M, v\models \chi_c^y$. As $y$ is either identical to $x$ or $x$ is not free in $\chi$, $\chi_c^y$ is the same formula as $\chi{_x^y}{_c^x}$ and $I_v(c)=[c]$, so by the substitution lemma $M, v_{[c]}^x\models \psi$ and $M, v_{[c]}^x\models \chi_x^y$. Furthermore, for all $[t]\in D$, if $M, v\models \chi_t^y$, then $M, v\models c=t$, i.e. $I_v(t)=I_v(c)$, i.e. $I_v(t)=[c]$. Let $v'$ be a $y$-variant of $v_{[c]}^x$, i.e. $v'=v{_{[c]}^x}{_{[s]}^y}$, for some $[s]\in D$. Either $y$ is identical to $x$ or $x$ is not free in $\chi$, so $v{_{[c]}^x}{_{[s]}^y}$ and $v$ agree on the assignments of elements of $D$ to all variables in $\chi$ except possibly $y$, and so $M, v{_{[c]}^x}{_{[s]}^y}\models \chi$ iff $M, v_{[s]}^y\models \chi$. So suppose now $M, v'\models \chi$ and $v'(y)\not=[c]$. $v'(y)=[s]$, so $[c]\not=[s]$. Then $M, v_{[s]}^y\models \chi$, and also if $M, v\models \chi_s^y$, then $M, v\models c=s$, i.e. $I_v(s)=I_v(c)$, i.e. $I_v(s)=[c]$. But $I_v(s)=[s]$, so $I_v(s)\not=[c]$. Hence $M, v\nvDash \chi_s^y$, and so by the substitution lemma, $M, v_{[s]}^y\nvDash \chi$. Contradiction. 
\medskip 

\noindent Finally, restrict the language again to the language of $\Delta$: structure $M$ constructed from $\Delta^+$ satisfies $\Delta$. This completes the proof of Theorem \ref{satisfiability}. \qed \end{proof}

\begin{theorem}[Completeness for Sequents]
	If $\models \Gamma\Rightarrow \Delta$, then  $\vdash \Gamma\Rightarrow \Delta$. 
\end{theorem} 

\begin{proof} Let $\neg \Delta$ be the negation of all formulas in $\Delta$. If $\models \Gamma\Rightarrow \Delta$, then $\Gamma, \neg \Delta$ is not satisfiable. Hence by Theorem \ref{satisfiability} it is inconsistent, and as they are both finite, $\vdash\Gamma, \neg\Delta  \Rightarrow \bot$. Hence by the properties of negation $\vdash \Gamma\Rightarrow \Delta$. \qed \end{proof}

\begin{theorem}[Completeness for Sets]
	If $\Gamma\models A$, then $\Gamma\vdash A$. 
\end{theorem}

\begin{proof} Suppose $\Gamma\models A$. Then $\Gamma, \neg A$ is not satisfiable, hence by Theorem \ref{satisfiability} it is inconsistent and $\Gamma, \neg A\vdash \bot$. So for some finite $\Sigma\subseteq \Gamma, \neg A$, $\Sigma\Rightarrow \bot$. If $\neg A\in \Sigma$, then by the deductive properties of negation, $\Sigma-\{\neg A\}\Rightarrow A$, and as $\Sigma-\{\neg A \}$ is certain to be a subset of $\Gamma$, $\Gamma\vdash A$. If $\neg A\not\in \Sigma$, then $\Sigma\Rightarrow A$ by the properties of negation, and again $\Gamma\vdash A$. \qed \end{proof}

By theorem 1 and 7 we also obtain the (strong) completeness of HRL.

\section{Conclusion}

Summing up, {\bf RL} saves the essential features of the Russellian approach to definite descriptions. It avoids problems like the arbitrary restriction of axiom $R$ to predicate symbols and scoping difficulties. In the semantics it retains the reductionist Russellian flavour in the sense that DD are not characterised by an interpretation function, but instead they are treated as a case in the clauses of the forcing definition for lambda atoms. In this respect {\bf RL} is different from the approach provided by Fitting and Mendelsohn \cite{FitMen98} which is closer to the Fregean tradition.

The rules of GRL are in principle direct counterparts of the tableau rules from \cite{IndZaw2} but with two important exceptions. The tableau rule corresponding to $(=-)$ is not restricted to atomic formulas and the tableau rule corresponding to $(\imath_2\Rightarrow)$ is not branching. Its counterpart in sequent calculus would be:

\vspace{.13in}

$(\imath_2\Rightarrow')$ \ $\dfrac{\mbox{$b_1=b_2, \Gamma \!\!\Rightarrow \Delta $}}{\mbox{$(\lambda x\psi)\imath y\varphi, \varphi[y/b_1], \varphi[y/b_2], \Gamma \!\!\Rightarrow \Delta$}}$

\vspace{.13in}

\noindent Such a non-branching rule is certainly much better for proof search, but it is not possible to prove the cut elimination theorem in its presence. The same applies to $(=-)$ without restriction to atomic formulas. In both cases the occurrences of arbitrary formulas $\varphi$ in the antecedent of the conclusion can be cut formulas and, in case the cut formula in the left premiss of the cut application is principal, it is not possible to make a reduction of the complexity of the cut formulas.

There is an interesting advantage of introducing the sequent characterisation of {\bf RL} over tableau formalisation from \cite{IndZaw2}. Since no rule specific to GRL has more than one active formula in the succedent they are also correct in the setting of intuitionistic logic as characterised by G1i \cite{tro:bas96}. It is sufficient to change the background calculus for the intuitionistic version (with $(\leftrightarrow\Rightarrow)$, $(\Rightarrow\vee)$ split into two rules, and  $(\Rightarrow C), (\Rightarrow W)$ deleted) and check that all proofs from section 3, 4 hold also for a (syntactically characterised) intuitionistic version of {\bf RL}. By comparison, the changes in the tableau setting would be rather more involved and connected with the introduction of labels for naming the states of knowledge in the constructed model.

The approach provided here may be modified also to cover some more expressive logics (like modal ones) and some other theories of DD like those proposed in the context of free logics. Some preliminary work in this direction is found in \cite{Indrzejczak2020a} and \cite{Indrzejczak2020b}. On the other hand the problems briefly mentioned in section 1 need serious examination and this may be carried out only after the implementation of the presented formal systems. This is one of the most important future tasks.

\paragraph{Acknowledgements.} We would like to thank Micha\l \ Zawidzki for his comments and suggestions.


\end{document}